\documentclass[a4paper,fleqn,usenatbib]{aa}

\usepackage[T1]{fontenc}
\usepackage{ae,aecompl}
\usepackage{graphicx}
\usepackage{amsmath}
\usepackage{amssymb}
\usepackage{color}
\usepackage{tabularx}
\usepackage{multirow}
\usepackage[format=hang ,justification=justified,font={footnotesize}]{caption}
\usepackage{varwidth}
\usepackage{epstopdf}
\usepackage{soul}
\usepackage{fixmath}
\usepackage[hang]{subfigure}
\usepackage{multirow}
\usepackage{multicol}
\usepackage{cuted}
\usepackage{hyperref}
\hypersetup{
    colorlinks=true,
    linkcolor=blue,
    filecolor=magenta,      
    urlcolor=blue,
    citecolor=blue,
}

\newcommand{\be}{\begin{equation}}
\newcommand{\ee}{\end{equation}}
\newcolumntype{$}{>{\global\let\currentrowstyle\relax}}
\newcolumntype{^}{>{\currentrowstyle}}

\newcommand{\planck}{\textit{Planck} }

\begin{document}

\title{Detection likelihood of cluster-induced CMB polarization}

\author{M. Mirmelstein\inst{1,2}\thanks{m.mirmelstein@sussex.ac.uk}, M. Shimon\inst{2}\thanks{meirs@tauex.tau.ac.il} \and Y. Rephaeli\inst{2,3}\thanks{yoelr@tauex.tau.ac.il}}

\institute{$^{1}$ Department of Physics \& Astronomy, University of Sussex, Brighton BN1 9QH, UK\\
$^{2}$ School of Physics and Astronomy, Tel Aviv University, Tel Aviv 69978, Israel\\
$^{3}$ Center for Astrophysics and Space Sciences, University of California, San Diego, La Jolla, CA, 92093, USA
}

\date{}

\abstract{
Nearby galaxy clusters can potentially induce sub-microkelvin polarization signals in the cosmic microwave background (CMB) at characteristic scales of a few arcminutes. We explore four such polarization signals induced in a rich nearby fiducial cluster and calculate the likelihood of their detection by a telescope project with capabilities such as those of the Simons Observatory (SO). In our feasibility analysis, we include instrumental noise, primordial CMB anisotropy, statistical thermal Sunyaev-Zeldovich (SZ) cluster signal, and point source confusion, assuming a few percent of the nominal telescope observation time of an SO-like project. Our analysis indicates that the thermal SZ intensity can be sensitively mapped in rich nearby clusters and that the kinematic SZ intensity can be measured with high statistical significance toward a fast moving nearby cluster. The detection of polarized SZ signals will be quite challenging but could still be feasible toward several very rich nearby clusters with very high SZ intensity. The polarized SZ signal from a sample of $\sim 20$ clusters can be statistically detected at $S/N \sim 3$, if observed for several months.}

\maketitle

\section{Introduction}

A wealth of information on the temperature anisotropy of the cosmic microwave background (CMB) and its E-mode polarization has been accumulated over the past two decades by multiple ground-based and balloon-borne telescopes as well as the WMAP and \planck satellites. In particular, \planck cosmic-variance-limited (CVL) measurements have mapped the temperature anisotropy down to $\sim 4'$ scales, and the South Pole Telescope (SPT) and the Atacama Cosmology Telescope (ACT) have already mapped the CMB temperature across a sky patch down to sub-arcminute resolution (\citealt{Akrami:2019bkn}, \citealt{Chown:2018fdj}, \citealt{Naess:2020wgi}).

Several ongoing ground-based CMB experiments, for example the upgraded versions of SPT and ACT -- SPTpol and ACTpol -- have detected CMB polarization on arcminute scales (e.g., \citealt{0004-637X-807-2-151}, \citealt{2017JCAP...06..031L}). The ongoing ground-based CMB experiment POLARBEAR (\citealt{adachi2020measurement}) and the forthcoming Simons Observatory (SO; \citealt{Ade:2018sbj}) will measure the CMB polarization to unprecedented precision on scales $\gtrsim 3'$, making the detection of polarization signals produced in gravitationally bound systems potentially feasible.

Temperature anisotropy in directions of galaxy clusters is induced by the scattering of the CMB off moving electrons; random and ordered motions give rise to the thermal and (the typically weaker) kinematic components of the Sunyaev-Zeldovich (SZ) effect, respectively. As is well established, measurements of the SZ effect constitute a powerful diagnostic tool that can yield important information on cosmological and cluster parameters (e.g., \citealt{refId0}, \citealt{0004-637X-832-1-95}, \citealt{2016MNRAS.461..248S}, \citealt{Ade:2018sbj}). Additional information can be extracted from the polarization state of the scattered radiation, which probes other combinations of cosmological and cluster parameters.

A prerequisite for cluster-induced polarization is the presence of a non-vanishing quadrupole in the incoming radiation when viewed in the scattering electron rest frame; the CMB is polarized upon Compton scattering off intracluster (IC) electrons. Since polarized Compton scattering blocks all moments other than the quadrupole at the rest frame of the scattering electron, CMB polarization toward individual galaxy clusters is a pristine measurement of the local quadrupole moment in the CMB. This local quadrupole could be of primordial origin or, alternatively, could yield information on IC gas properties, that is, the cluster geometry, its peculiar velocity in the Hubble frame, gas temperature, or a combination thereof. Two such quadrupole sources, linear in the optical depth, are explored here: the primordial CMB quadrupole and quadrupole anisotropy associated with the transversal second-order Doppler component of the bulk motion of the cluster.

Two other relevant polarization components arise from double scattering in the same cluster (\citealt{1980MNRAS.190..413S}, \citealt{1999MNRAS.310..765S}). First, scattering induces temperature anisotropy either via the thermal or kinematic SZ effects. If this temperature anisotropy contains a local quadrupole moment, polarization is induced upon second scattering. For a typical Thomson optical depth of IC gas, $\tau \sim 0.01$, these $\tau^2$-dependent components are clearly very weak, with the thermal effect sourcing the larger of the two. 

Cluster polarization signals could possibly provide a way to increase the sampling of the primordial quadrupole by the additional linear polarization it induces, thereby lowering the statistical uncertainty of measurement of this quadrupole moment (\citealt{1997PhRvD..56.4511K}, \citealt{2006PhRvD..73l3517B}), a possibility that was further elaborated upon by, for example, \citet{2014PhRvD..90f3518H}, \citet{Yasini:2016pby}, and \citet{Louis:2017hoh}. It has also been suggested that the CMB quadrupole anisotropy deduced from SZ polarization maps could possibly enable the reconstruction of the cosmological reionization history (\citealt{PhysRevD.97.103505}).

A quantitative study of cluster-induced polarization was begun by \citet{1980MNRAS.190..413S} and continued with the works of \citet{1999MNRAS.310..765S} and \citet{1999MNRAS.305L..27A}. Polarization levels and spatial distribution were determined in a detailed analysis of a cluster simulated with the hydrodynamical {\it Enzo} code by \citet{2006MNRAS.368..511S}. Power spectra of the statistical (all-sky) cluster-produced polarization, which is clearly much smaller than the primordial CMB polarization on the relevant angular scales, was also studied (e.g., \citealt{Cooray:2004cd}, \citealt{2006MNRAS.368..511S}).

In order to assess the likelihood of detection of the typically weak polarization signals from possible cluster targets of current and anticipated very high resolution ground-based polarization-sensitive experiments, a realistic feasibility analysis is needed. The main objective of the work presented here is to carry out detailed estimates of cluster polarization signals and to assess their detectability toward several rich nearby clusters by a project such as the SO, which will clearly be able to map the much larger (total) thermal SZ intensity, as well as the kinematic SZ signal from clusters with high ($\gtrsim 1000$ km s$^{-1}$) radial velocity.

\section{Polarization induced by scattering in clusters}
\label{sec:polarization_induced_by_scattering in_clusters}

The scattering of the CMB by (hot) IC gas changes its spectro-spatial distribution in the direction of the cluster. The change in the spectral brightness temperature (expressed in terms of the spectral intensity $I_{\nu}$)
\be
T(\nu)=\frac{h\nu}{k_B\ln(1+\frac{2h\nu^{3}}{c^{2}I_{\nu}})},
\ee
simplified in the (most) relevant Rayleigh-Jeans limit to $T(\nu) \simeq \frac{c^2}{2k_B\nu^2}I_{\nu}$ (where the physical constants $c$, $h$, and $k_{B}$ have their usual meaning), is expressed by the fractional temperature change of the thermal SZ effect (\citealt{1972CoASP...4..173S})
\be
\frac{\Delta T}{T}=yf(x),
\ee
where
\be
\begin{split}
&f(x) = x\coth(x/2)-4, \\
&y = \int\sigma_{T} n_{e}\frac{k_BT_{e}}{m_{e}c^{2}}dl.
\label{eq:f_x_def}
\end{split}
\ee
Here, $x=h\nu/(k_BT_{\gamma})$ is the dimensionless frequency expressed in terms of the CMB temperature, $T_{\gamma}$; $y$ is the Comptonization parameter; $\sigma_T$ is the Thomson cross section; $m_{e}$, $n_{e}$, and $T_{e}$ are the electron mass, number density, and temperature, respectively; and the integration is along the line of sight (los). We denote the dimensionless gas temperature by $\Theta\equiv k_BT_{e}/(m_{e}c^{2})$.

The typically smaller kinematic SZ effect is proportional to the los velocity component of the cluster, $v_{r}$; since it is a first-order Doppler shift of the CMB, the temperature change it induces is frequency-independent,
\be
\frac{\Delta T}{T}=-\int\sigma_{T}n_{e}\beta_{r}dl,
\ee
with $\beta_{r}=\frac{v_{r}}{c}$.
Compton scattering can polarize incident radiation if it has a quadrupole moment in the rest frame of the scattering electron; the quadrupole is the only multipole transmitted by scattering (assuming photons are soft in the electron rest frame, which is virtually always the case for the thermal electron populations considered here). The incident CMB radiation has a global quadrupole moment in addition to a small non-vanishing quadrupole moment ($O(10^{-5})$), which is induced by (first) scattering in the cluster (i.e., the SZ effect). The degree of linear polarization and its orientation are determined by the two Stokes parameters
\be
\begin{split}
Q=&\frac{3\sigma_{T}}{16\pi}\int n_{e} dl \int \sin^{2}\theta\cos
(2\phi) T(\theta,\phi) d\Omega, \\
U=&\frac{3\sigma_{T}}{16\pi}\int n_{e} dl\int \sin^{2}\theta\sin
(2\phi) T(\theta,\phi) d\Omega,
\label{stokes_params}
\end{split}
\ee
where $\theta$ and $\phi$ define the directions of the incoming photon relative to the outgoing direction (taken to be the z-axis), with $\phi$ measured from the x-axis. The Stokes parameters are defined with respect to the x- and y-axes; $d\Omega$ is an element of integration over the solid angle, and $T(\theta,\phi)$ is the temperature of the incident radiation. Since the los is taken to be along the z-axis for convenience, the angles $\theta$ and $\phi$ are actually defined with respect to the outgoing photon in this system. The average electric field defines the polarization plane with a direction given by
\be
\alpha=\frac{1}{2}\tan^{-1}\left(\frac{U}{Q}\right),
\ee
and the total polarization is defined as
\be
P\equiv\sqrt{Q^{2}+U^{2}}.
\ee

In the following, we describe the polarization generated by Compton scattering when a quadrupole moment is induced either by electrons moving at the cluster radial velocity or electrons randomly moving in the hot gas with a velocity dispersion that is characterized by the gas temperature $T_{e}$.

\subsection{Polarization of isotropic incident radiation}

Scattering of the CMB in a cluster at rest in the CMB frame results in local anisotropy due to the different path-lengths of photons arriving from various directions to a given point (e.g., \citealt{2014MNRAS.437...67C}). This anisotropy provides the requisite quadrupole moment; second scatterings then polarize the radiation (\citealt{1980MNRAS.190..413S}, \citealt{1999MNRAS.310..765S}). By symmetry, the measured net polarization (essentially) vanishes if the cluster is not resolved. Nonetheless, and since we consider here the performance of an arcminute-resolution experiment, it is useful to explore the signal associated with double scattering since it is {\it a priori} expected to dominate over the other polarization signals in rich clusters (\citealt{2006MNRAS.368..511S}), for which $\tau^{2}$ is not negligibly small.

The CMB appears anisotropic in the frame of a non-radially moving cluster; consequently, scattering by IC electrons polarizes the radiation. Two polarization components are induced: The first is linear in the cluster velocity component transverse to the los, $v_t \equiv\beta_{t}c$, but quadratic in $\tau$; the second is linear in $\tau$ but quadratic in $\beta_{t}$. The quadrupole required for the latter is induced by the (transversal) second-order Doppler effect. The spatial patterns of the various polarization components can be readily determined when the gas distribution is spherically symmetric, in the "hard sphere" approximation for gas density and temperature (e.g., \citealt{1980MNRAS.190..413S}, \citealt{1999MNRAS.310..765S}). More realistic gas distributions, substructure, and high internal velocities result in complicated polarization patterns (\citealt{2004MNRAS.347..729L}, \citealt{2006MNRAS.368..511S}).

\subsubsection{The kinematic polarization components}

The degree of polarization induced by double scattering in a non-radially moving cluster was determined by \citet{1980MNRAS.190..413S} in the simple case of uniform gas density,
\be
P=\frac{1}{40}\tau^{2}\beta_{t}.
\ee
This polarization component is frequency-independent in equivalent temperature units since the requisite quadrupole is generated by the kinematic SZ effect, a first-order Doppler shift. A more complete calculation of this and the other polarization components was carried out by \citet{1999MNRAS.310..765S}. Viewed along a direction $\hat{\bf{n}}=(\theta,\phi)$, the temperature anisotropy at a point $(X,Y,Z)$, $\Delta T(X,Y,Z,\theta,\phi)$, leads to polarization upon second scattering. The Stokes parameters are calculated from Eq.~\ref{stokes_params},
\be
\begin{split}
\left(Q\pm iU\right)(X,Y)=&\frac{3\sigma_{T}}{16\pi}\int dZ n_{e}(X,Y,Z) \\ &\times \int d\Omega\sin^{2}(\theta)e^{\pm i2\phi}\Delta T(X,Y,Z,\theta,\phi)\\
=&\frac{3\sigma_{T}}{4\pi}\int dZ n_{e}(X,Y,Z) \\ &\times \int d\Omega Y_{2}^{\pm 2}(\theta,\phi)\Delta T(X,Y,Z,\theta,\phi),
\label{eq:pol_kinematic}
\end{split}
\ee
where $Y_{2}^{\pm 2}$ are the $\ell=2$ and $m=\pm 2$ spherical harmonics, and we recall that $\theta$ and $\phi$ are defined by a system centered at the scattering electron. The temperature change resulting from first scatterings is
\be
\begin{split}
\frac{\Delta T(X,Y,Z,\theta,\phi)}{T}=&\sigma_{T}
\int d\vec{l}(X',Y',Z',\theta,\phi) \\&\times n_{e}(X',Y',Z',\theta,\phi)\hat{n}\cdot
\mathbold{\beta}(X',Y',Z'),
\label{eq:temp_kinematic}
\end{split}
\ee
and the optical depth through the point $(X,Y,Z)$ in the direction
$(\theta,\phi)$ is
\be
\tau(X,Y,Z,\theta,\phi)=\sigma_{T}\int n_{e}(X',Y',Z')d\vec{l}
(X',Y',Z',\theta,\phi).
\ee
Thus, $Q(X,Y)$ and $U(X,Y)$ fully describe the linear polarization field.

The second kinematic polarization component is $\propto \tau\beta_{t}^{2}$; it is generated by virtue of the fact that the radiation appears anisotropic in the electron frame if the electron motion has a non-vanishing transverse component. The polarization of singly scattered radiation is then calculated (\citealt{1999MNRAS.310..765S}) by using Eq.~\ref{stokes_params}. The Stokes parameters for this polarization component are
\be
\begin{split}
Q=&\frac{1}{20}\tau(\beta_{x}^{2}-\beta_{y}^{2})g(x), \\
U=&\frac{1}{10}\tau\beta_{x}\beta_{y}g(x),
\end{split}
\ee
where
\be
g(x)\equiv x\coth\frac{x}{2}.
\label{g_x_def}
\ee
By applying a rotation in the sky (x-y) plane (i.e., working in a frame with the x-axis aligned with the transverse cluster velocity), it can be shown that $Q=\frac{1}{20}\tau(\beta_{x}^{2}+\beta_{y}^{2})g(x)$ and $U=0$.

The polarization direction in our arbitrary coordinate system is
\be
\alpha=\frac{1}{2}\tan^{-1}\left(\frac{2\beta_{x}\beta_{y}}{\beta_{x}^{2}-\beta_{y}^{2}}\right)
.\ee

\subsubsection{The thermal polarization component}

As with the $\tau^{2}\beta_{t}$ component discussed above, double scattering off electrons moving with random thermal velocities can induce polarization that is proportional to $\tau^{2}\Theta$. The anisotropy introduced by single scattering is the thermal component of the SZ effect with temperature change $\Delta T_{t}$. Its dependence on frequency is quantified by the spectral function $f(x)$ as defined in Eq.~\ref{eq:f_x_def}. In particular, this spectral function changes signs at the crossover frequency $x_{0}=3.83$; consequently, $Q$ and $U$ flip signs, and the polarization pattern locally rotates on the sky by $90^{\circ}$. Globally, this results in the polarization pattern changing from radial at $x<x_{0}$ to tangential at $x>x_{0}$ (\citealt{1999MNRAS.310..765S}). The calculation of the effect follows Eqs.~\ref{eq:pol_kinematic} and~\ref{eq:temp_kinematic}, but with the latter replaced by an integral of the pressure, as appropriate for the thermal SZ effect induced by the first scattering.

\subsection{Polarization of anisotropic incident radiation}

The large-scale anisotropy of the CMB includes a global quadrupole moment at the level of $\approx 15\;\mu K$, as measured by the all-sky surveys COBE, WMAP, and \planck (\citealt{refId02}, and references therein). Knowing that the probability for generating polarization by scattering in clusters is a fraction $\sim\frac{\tau}{10}$ of the incident quadrupole, and that $\tau \sim 0.01$, we expect the resulting polarization signal to be $\approx$ 15 nK, tiny by all measures, including the currently projected high sensitivity levels. A study of the detectability prospects of this effect has concluded that this component could be marginally detected at a $\sim 2\sigma$ confidence level from a joint analysis of polarization maps of 550 clusters (\citealt{2014PhRvD..90f3518H}). It seems reasonable to expect that other cluster-induced CMB polarization signals and source confusion will likely degrade this projected detection level.

\section{Detectability of the polarization signal}

The main objective of this work is to determine values of the signal-to-noise ratios (S/N) for the detection of polarized CMB radiation induced by a rich cluster. Assuming that temperature and two polarization (Q and U Stokes parameters, or the equivalent E- and B-modes) maps of the cluster are available, one can use their auto- and cross-correlations to estimate the S/N for cluster detection. The cluster is placed at various redshifts in the interval $[0.02, 0.2]$, and its gas density and temperature are described by a polytropic equation of state with a $\beta$-profile.

Polarization maps and polarization levels for the cluster are produced for all components discussed above: $\propto\tau Q_{\rm prim}$, $\propto\tau\beta_{t}^{2}$, $\propto\tau^{2}\Theta$, and $\propto\tau^{2}\beta_{t}$. These have different characteristic spatial and spectral signatures, and therefore the feasibility of their measurement will obviously depend on the telescope sensitivity and angular resolution. Since ground-based experiments normally observe through only a few narrow atmospheric spectral bands, the main difference between the projected S/N achievable with the different telescopes will largely depend on their nominal instrumental noise levels. Our analysis does not account for beam systematics as our goal is to assess the expected nominal S/N independently of instrument-specific modeling uncertainties. This source of systematics is important for making inferences regarding the primary polarization signal, which is statistically isotropic; ideally, this would not be the case for a cluster when the scanning strategy of the telescope is optimized for such an observation. Typically, beam systematics are dominant on sub-beam scales with appreciable leakage from total-to-polarized intensity, and therefore our estimates would in this sense provide over-idealized bounds on the expected polarization signal. (We do account for point source confusion, as specified below.)
It should also be noted that since beam systematics virtually only affect cluster-induced polarization, it is expected that cross-correlation with the kinematic or thermal effects could be used in filtering out any such significant contamination. Null-hypothesis tests applied to "quiet" sky patches with no known clusters could also be used to monitor large systematics.

Polarization maps are decomposed into their E- and B-modes. This is required because the primordial CMB contribution is typically given in terms of the pure-parity E- and B-modes. For the same definite-parity property of the E- and B-modes, it is advantageous to work on this basis as part of the validation of our numerical calculation: A trivial radially independent, circularly symmetric signal is expected to generate no helicity-preferred polarization, and therefore the residual B-mode due to numerical errors would ideally be significantly suppressed compared to E-mode polarization.

The maps were obtained from the convolution of the simulated polarization maps with a 1' Gaussian beam and at $\nu=150$ GHz, the frequency band that is common to most current ground-based CMB experiments. The maps can be scaled to other frequencies using the spectral functions $f(x)$ and $g(x)$ from Eqs.~\ref{eq:f_x_def} and~\ref{g_x_def}, respectively. The S/N levels were determined as a function of redshift on the observed sky patch, here taken to be a square for simplicity. The sky patch is assumed to be the field of view (FOV), the minimal sky patch that can be observed by a given telescope. We assume that the telescope is locked on the same target for $1\%$ and $4\%$ of its nominal observing time. For reference we also report results for no instrumental noise, which we refer to as a CVL experiment, with all astrophysical and cosmological foregrounds left unchanged. We used the nominal telescope sensitivity $\Delta_{T}$ (in units of microkelvin-arcminutes) and scale it according to $\Delta_{T}\propto\sqrt{\Omega/t}$, where $\Omega$ is the solid angle subtended by the observed sky patch and $t$ is the observation time. The SO-like telescope specifications adopted in this work are shown in Table~\ref{table:specs}. In simulating cluster observations, we assumed a square FOV, 7.8$^{\circ}$ per side as in \cite{Ade:2018sbj}. Our CVL results bracket possible improvement in S/N due to ideal sky patch choice that will be allowed once the FOV can reach a significantly smaller size than the nominal value assumed in our current analysis.

Throughout, we adopted the flat-sky approximation, which is highly accurate on galaxy cluster scales but is much less accurate at the angular size of the full FOV. However, the power associated with these large-scale modes is negligible as they are outnumbered by the small-scale modes, and the power that they contribute is outweighed by the small-scale power. The sub-percent error resulting from this approximation is negligible compared to other modeling uncertainties. Clearly, the main contribution to the cluster (temperature and polarization) maps are at sufficiently large multipoles, $\ell$, which correlate with typical spatial SZ features on arcminute scales (e.g., \citealt{2004NewA....9..159S}). At higher multipoles, beam dilution suppresses any information from sub-structures in the cluster map; this happens before the minimum angular scale, defined by the simulation resolution, is reached. For the expected noise power spectrum, $C_{\ell}^{XY}$ with $X, Y\in\{T, E, B\}$, the main contributions are the primordial CMB temperature anisotropy and polarization, detector noise, integrated SZ (temperature only), and point sources.

Assuming certain linear Wiener-type filtering is applied to the observed T, E, and B maps which down-weights modes according to their respective S/N (per mode, derived from their spectro-spatial dependence), the expected S/N for the detection of a specific intensity or polarization component from using all possible auto- and cross-correlations and all observed modes at all spectral bands is (e.g., \citealt{1996MNRAS.281.1297T}, \citealt{2002MNRAS.336.1057H}, \citealt{2010MNRAS.403.2120L})
\be
\left(\frac{S}{N}\right)^2 = \sum\limits_{\nu_1,\nu_2}\int\frac{d^2\ell}{(2\pi)^2}{\bf S_{\ell}}(\nu_1){\bf N}_{\ell}(\nu_1,\nu_2)^{-1}{{\bf S}}_{\ell}(\nu_2)^\dagger,
\label{eq:S2N}
\ee
where $\nu_1\leq\nu_2$, ${\bf S_{\ell}}=(T_{{\bf \ell}},E_{{\bf \ell}},B_{{\bf \ell}}),$ 
 $X_{\ell}$ is the Fourier transform of real-space map $X$, and ${\bf N}_{\ell}^{-1}$ is the inverse of the covariance matrix,
\be
{\bf N}_{\ell} = \left(\begin{array}{c c c}
C_{\ell}^{TT} & C_{\ell}^{TE} & C_{\ell}^{TB} \\
C_{\ell}^{TE} & C_{\ell}^{EE} & C_{\ell}^{EB} \\
C_{\ell}^{TB} & C_{\ell}^{EB} & C_{\ell}^{BB} \\
\end{array}\right).
\label{clmatrix}
\ee
While the spectral dependence of the target signals is theoretically known, their spatial dependence depends on the cluster density, temperature, and velocity profiles. The impact of model uncertainties is discussed below.

Overall, there are six possible pairings: TT, EE, BB, TE, TB, and EB. Here ${\bf N}_{\ell}$ should be understood as a 3x3 symmetric matrix per each multipole whose XY component is the power spectrum $C_{\ell}^{XY}$ at that $\ell$. The $C_{\ell}^{TB, \rm{prim}}$ and $C_{\ell}^{EB, \rm{prim}}$ vanish in the standard cosmological model. The detector noise $C_{\ell}^{XY, \rm{det}}$ ideally vanishes whenever $X\neq Y$, and the all-sky integrated polarization from the entire cluster population is negligible (\citealt{2003NewAR..47..839B}, \citealt{2006MNRAS.368..511S}), and, as such, so is the expected EB-TB correlation. Since the temperature-polarization correlation of point sources is not known, we assumed that all cross-correlation power spectra of the noise contributions, with the exception of $C_{\ell}^{TE}$, are negligible. 
Then, Eq.~\ref{eq:S2N} significantly simplifies to
\be
\begin{split}\label{eq:s2n_T_P}
&\left(\frac{S}{N}\right)_{T+P}^{2}=\sum\limits_{\nu_1,\nu_2}\int \frac{d^{2}{\ell}}{(2\pi)^{2}}\times
\\ &\Bigg[\frac{T_{{\ell},\nu_1}T_{{\ell},\nu_2}^*C_{\ell}^{EE}-2{\rm Re}(T_{{\ell},(\nu_1}^{*}E_{{\ell},\nu_2)})
C_{\ell}^{TE}+E_{{\ell},\nu_1}E_{{\ell},\nu_2}^*C_{\ell}^{TT}}
{C_{\ell}^{TT}C_{\ell}^{EE}-(C_{\ell}^{TE})^{2}}
\\ &+\frac{B_{{\ell},\nu_1}B_{{\ell},\nu_2}^*}{C_{\ell}^{BB}}\Bigg],
\end{split}
\ee
where the various $C_{\ell}$ are the (generally frequency-dependent $C_{\ell}\equiv C_{\ell}(\nu_1,\nu_2)$) power spectra characterizing the various noise contributions to both temperature and polarization. In practice, these noise power spectra are obtained from the map itself -- here we only adopt the expected noise power spectra; for example, for the primordial CMB, we used the output of the Boltzmann code CAMB (\citealt{2000ApJ...538..473L}). We assumed a detector noise level compatible with that of the SO-like experiment (\citealt{Ade:2018sbj}), and, from recent high resolution observations by SPTpol and ACTpol, we estimated the noise level due to polarized and unpolarized point sources (\citealt{2011ApJ...743...28K}, \citealt{Crites:2014prc}). In addition, in the case of temperature measurements, we added the all-sky statistical SZ noise contribution to temperature anisotropy as calculated by our numerical SZ code (\citealt{2009MNRAS.399.2088S}). It is shown in this last paper and in \citealt{Cooray:2004cd} that the all-sky cluster-induced polarization noise is orders of magnitude smaller than, for example, the primordial polarization signal, and it is therefore neglected in the present analysis. Given $Q$ and $U$ maps on the flat sky, the E- and B-modes are defined as the Fourier modes of the polarization field $P\equiv Q\pm i U$ in a spin-2 plane wave expansion
\be\label{eq:QUEB}
(Q\pm iU)({\bf \hat n})\equiv\int\frac{d^{2}{\ell}}{(2\pi)^{2}}
\left[E_{{\boldsymbol\ell}}\pm iB_{{\boldsymbol\ell}}\right]
e^{\pm i2\phi_{\ell}}e^{i{\boldsymbol\ell}\cdot\mathbold{\theta}}
,\ee
where $\mathbold{\theta}$ is the 2D angle vector.

To estimate the polarization-only S/N, setting $T_{\ell}=0$ in Eq.~\ref{eq:s2n_T_P} results in 
\be
\begin{split}\label{eq:s2n_P}
\left(\frac{S}{N}\right)_{P}^{2} =& \sum\limits_{\nu_1,\nu_2}\int \frac{d^{2}{\ell}}{(2\pi)^{2}}\times
\\ &
\Bigg[\frac{E_{{\ell},\nu_1}E_{{\ell},\nu_2}^*C_{\ell}^{TT}}
{C_{\ell}^{TT}C_{\ell}^{EE}-(C_{\ell}^{TE})^{2}} +\frac{B_{{\ell},\nu_1}B_{{\ell},\nu_2}^*}{C_{\ell}^{BB}}\Bigg].
\end{split}
\ee
By symmetry, it is clear that the $\propto\tau^{2}\Theta$ polarization signal (\citealt{1999MNRAS.310..765S}) contributes purely to the E-mode. It is easy to see that for all other individual signals explored in this work, with the exception of the $\propto\tau^{2}\beta_{t}$ signal, the U Stokes parameter could be made to vanish by a specific choice of reference frame on the sky. In this case, as is evident from Eq.~\ref{eq:QUEB}, we expect comparable contributions from the E- and B-modes. Therefore, in the case of detector-noise dominated observations, we expect comparable contributions by the E- and B-modes to the overall S/N. However, in the event of sufficiently deep observations, as assumed here, the noisier E-mode has a relatively small contribution to the S/N. For example, the primordial CMB spectra satisfy $C_{\ell}^{EE} - {C_{\ell}^{TE}}^2/C_{\ell}^{TT} \approx C_{\ell}^{EE}>C_{\ell}^{BB}$ over the relevant multipole range.
Due to the relative weight of the B-modes in our analysis, it is of interest to estimate the amount of leakage that results from the E-B decomposition across the FOV. We estimate this effect in the following section.

\section{Results}

We calculated the expected S/N for both the total and polarized SZ intensities in a rich cluster with a total mass of $M_{\rm tot}=2\times10^{15}M_{\odot}$ at an observed redshift in the range $0.02 \leq z \leq 0.2$. The IC gas is described by a polytropic equation of state
\be
P_g = P_{g,0} \Big(\frac{\rho_g}{\rho_{g,0}}\Big)^{\Gamma},
\ee
where $P_g$ is the gas pressure, $P_{g,0}$ is the central pressure, $\rho_g$ is the gas density, $\rho_{g,0}$ is the central gas density, and $\Gamma$ is the adiabatic index (which is related to the polytropic index $n$ through $\Gamma=1+1/n$). We adopted the commonly used value $\Gamma=1.2$ ($n=5$). A $\beta$-profile (\citealt{1978A&A....70..677C}) is taken for the gas spatial distribution
\be
\rho_g = \rho_{g,0} \Bigg[1+\Big(\frac{r}{r_c}\Big)^2\Bigg]^{-\frac{3}{2}\beta}
,\ee
where $r_c$ is the (gas) core radius; typical values of $r_c$ and $\beta$ (deduced from measurements of the cluster X-ray surface brightness) are $r_c = 200$ kpc and $2/3$, respectively. We constructed a 3D semi-analytic simulation of the gas distribution with $\sim 200$ cubical cells with a minimum ("resolution") side of $20$ kpc; since the cluster virial radius (determined in terms of a mean cluster mass density of 200 times the critical density of the Universe at the cluster redshift) is $R_V = 2.2$ Mpc, it is fully contained in the simulation volume.

The gas temperature and its profile are determined from the expression for the (ideal gas) pressure $P_g = \rho_g k_B T_g/\mu$,
\be
T_g=T_{g,0}\Big({\frac{\rho_g}{\rho_{g,0}}}\Big)^\frac{1}{n},
\ee
where $\mu$ is the mean molecular weight ($\approx 0.6$ in a fully ionized cosmic gas) and $T_0$ is the central temperature whose fiducial value is set to $T_0=10^8$ K. The central density, 
\be
\rho_{g,0} = f_gM_{\rm tot} \Bigg[4\pi \int\limits_0^{R_V}r^2(1+(r/r_c)^2)^{-\frac{3}{2}\beta} dr \Bigg]^{-1},
\ee
is obtained from the total mass (within the virial radius) and the gas mass fraction $f_g$.
The electron number density is $n_e \simeq 6\rho_{g}/(7 m_{p})$. In our numerical estimates, we take $f_g=0.12$ (e.g., \citealt{2015MNRAS.450..896D}) and consider the cluster to be moving at a velocity of $1000~\text{km}~\text{s}^{-1}$, both along and across the los.

\begin{table}
\caption{Specifications used in modeling detector white noise for a SO-like experiment~(\citealt{Ade:2018sbj}). The sensitivity for intensity measurements is related to that of the polarization by $\Delta_P =\sqrt{2}\Delta_T$, and $f_{\rm sky}$ is the observed sky fraction.}
\label{table:specs}
\centering
\begin{tabular}{   c  c  c  c  c  c }
\hline
$f_{\rm sky}$                           &               $\nu$           & $\theta_{\text{\tiny FWHM}}$    &               $\Delta_{P}$                    \\
                                                                &               (GHz)         &                               (arcmin)                                        & ($\mu \text{K}$-arcmin) \\ \hline\hline
\multirow{3}{*}{0.4}    &                 95            &                                 1.6                                            &                       5.8                                     \\
                                                        &                150             &                                 1.0                                           &                         6.3                                     \\
                                                      &                  220             &                                 1.3                                           &                         15.0                                    \\ \hline
\end{tabular}
\end{table}

\begin{figure}
\begin{center}
\subfigure[Temperature.]{\includegraphics[width=5.5cm,trim={0  2.4cm 0 0}]{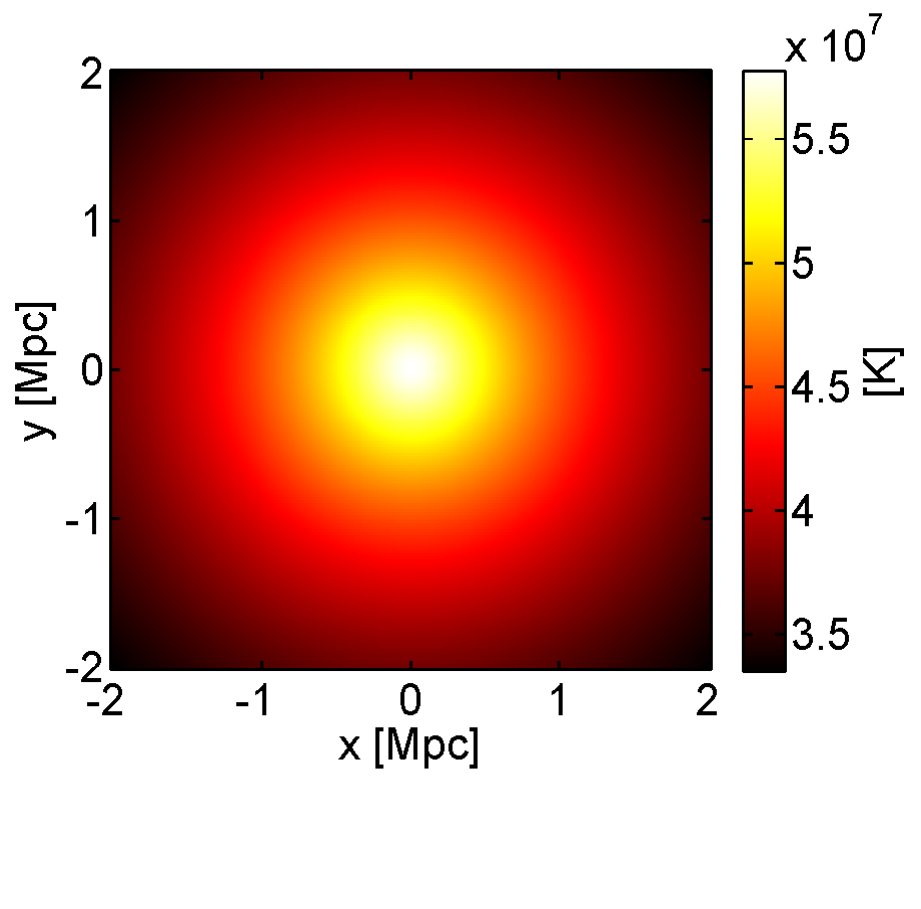} \label{fig:sim_T}}  \\ \vspace*{-4mm}
\subfigure[Electron number density.]{\includegraphics[width=5.5cm,trim={0  2.4cm 0 0}]{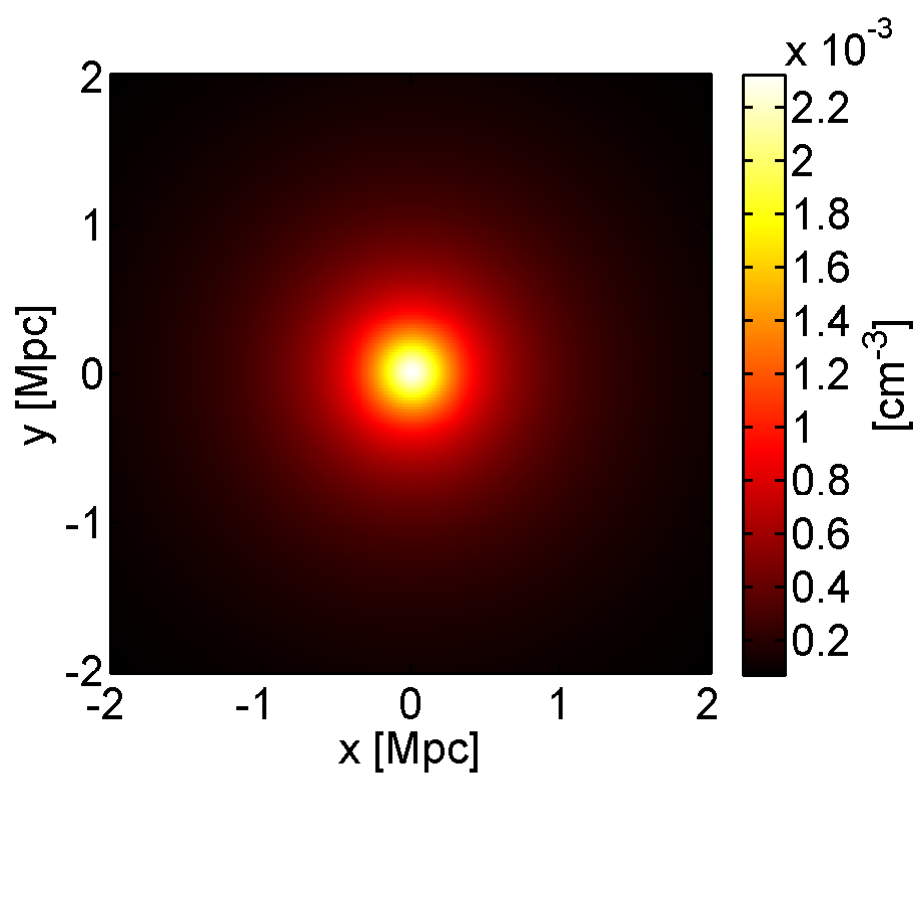} \label{fig:sim_n_e}}  \\ \vspace*{-4mm}
\subfigure[Optical depth.]{\includegraphics[width=5.5cm,trim={0  2.4cm 0 0}]{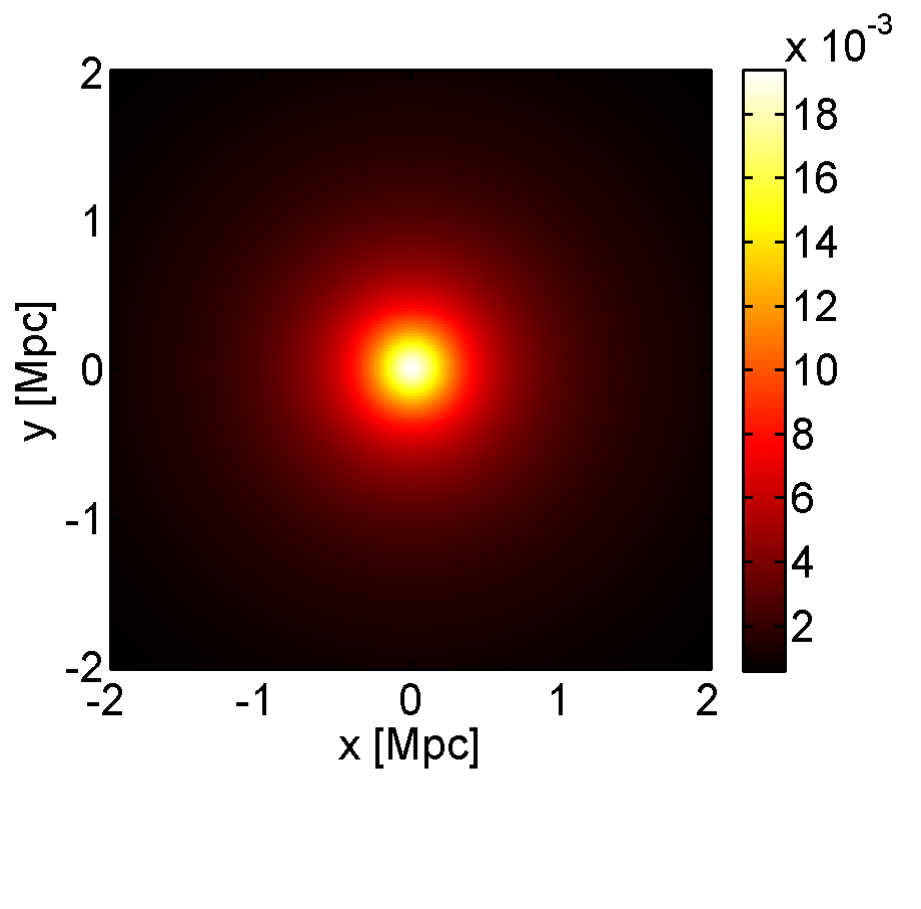} \label{fig:sim_tau}}
\end{center}
\caption{Maps of the projected~\subref{fig:sim_T} temperature,~\subref{fig:sim_n_e} electron number density, and~\subref{fig:sim_tau} optical depth in our model cluster. The electron number density -- and, consequently, the optical depth -- are more centrally concentrated than the temperature.}
\label{fig:sim}
\end{figure}

The projected temperature, gas density, and optical depth maps in our fiducial rich cluster are shown in Fig. \ref{fig:sim}. At 150 GHz, the thermal component is larger than the kinematic component by a factor of $\sim4$ and is somewhat more centrally concentrated due to its dependence on both temperature and density, whereas the kinematic component is $\propto\tau$ (Figure~\ref{fig:sim_tau}). Observing the cluster at a spectral band around $\nu_{c}\sim218$ GHz will result in a substantially smaller contribution from the former component due to its sharply decreasing profile close to $\nu_{c}$. The Q and U polarization maps are shown in Appendix~{\ref{appendix_A}} for the $\tau^2\beta_{t}$ and $\tau^2\Theta$ components.

In practice, applying the filters to the temperature anisotropy and polarization maps (that ultimately result in the S/N of Eq.~\ref{eq:s2n_P}) only requires an assumption about the 2D profiles of the signals. For example, in the case of the thermal SZ effect, this amounts to the profile of the Comptonization parameter. Knowing the profiles to a very high precision is not essential; rather, it is the characteristic angular scale of the 2D profile that determines the relevant scales over which the noise is down-weighted by the filter. Assuming the profile is perfectly known but the scale is not, it is possible to apply the same filter, each time with a different scale, until the highest S/N is achieved. We discuss below the impact of uncertainties in specific cluster model parameters on values of the S/N. As to the noise power spectra appearing in the denominator of Eq.~\ref{eq:s2n_P}, these are computed in real data analysis directly from the T, E, and B maps for the various frequency channels. This procedure is valid under the assumption that the sought-after signal does not significantly contribute to the computed power spectra; this assumption is clearly justified in our case, as is made evident below. In this work, and as is always the case when forecasting observational results from upcoming experiments, we simulate the expected noise power spectra as no maps are yet available from the forthcoming SO. Our noise budget includes the primordial CMB power spectra (computed with the Boltzmann-solver CAMB assuming the standard vanilla cosmological parameters), the statistical signature of the SZ effect contribution to temperature anisotropy noise, detector noise power spectra (Table~\ref{table:specs}), and point source power spectra.

\begin{figure*}
\includegraphics[width=\textwidth]{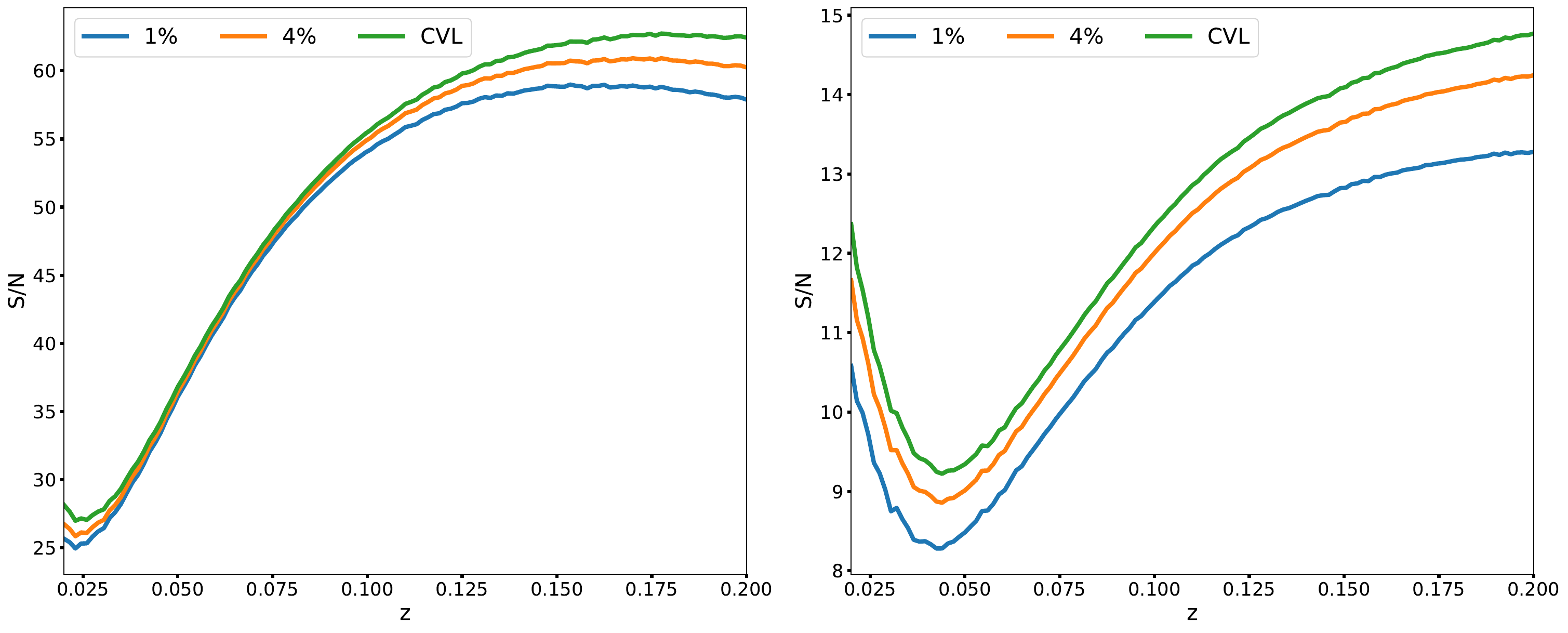}
\caption{Predicted S/N for measurement of the thermal (left) and kinematic (right) SZ components in temperature-equivalent units for the polytropic cluster simulation for an SO-like experiment with $1\%$ (blue) and $4\%$ (orange) observation time and CVL (green). Point source noise contamination is assumed at the deduced upper limit (specified in the text).}
\label{fig:S2N_Temperature}
\end{figure*}

The predicted S/N of both the total and polarized intensities include the confusion due to emission by (unclustered) radio and infrared galaxies. We adopt the currently deduced upper bounds on the power spectra ($C_{\ell}$) of point sources $13~\text{nK}^2$ (\citealt{2011ApJ...743...28K}) and $0.28~\text{nK}^2$ (\citealt{Crites:2014prc}) for total and polarized intensities, respectively. While the quoted values were given specifically for 150 GHz and with a multipole dependence, we treat them as constant in harmonic space and do not scale them with frequency as it is not immediately clear to us that a conservative upper bound deduced at a given frequency can be justifiably scaled to upper bounds at other frequencies, even if the frequency-dependence of the foreground is theoretically known.
Estimates of the S/N levels for the thermal SZ effect toward the model cluster, if observed for the specified integration times, are plotted as a function of redshift in the left-hand panel of Fig.~\ref{fig:S2N_Temperature}. In estimating the detection feasibility of the thermal SZ signal, the kinematic signal is considered as a source of noise, and, as such, it is included in the noise budget, thereby increasing the noise variance in Eq.~\ref{eq:S2N}. 
While the amplitude of the kinematic SZ signal from an individual cluster is unknown, the relevant quantity in the calculation of the S/N in Eq.~\ref{eq:S2N} is the power spectrum associated with the all-sky signal that is calculated using our numerical SZ code. The power spectra associated with the various cluster-induced polarization signals explored in the present work are much smaller than the primordial polarization power spectra (\citealt{2009MNRAS.399.2088S}).

Unsurprisingly, highly significant detection of the thermal SZ effect is predicted already at the lowest respective $1\%$ observation time. It is also evident that for an SO-like experiment, this observation time can yield detection at levels as high as those of the CVL. We expect that, with an additional high frequency band such as that which is planned for the SO, the overall S/N levels will, of course, be even higher. 

We also estimate the detection of the kinematic component when the thermal component is considered as a noise source, namely by applying a filter to the polarization maps that down-weights Fourier modes and frequency channels over which the thermal component dominates, as described above in Eq.~\ref{eq:S2N}. Estimated S/N levels in this case are shown in the right-hand panel of Fig.~\ref{fig:S2N_Temperature}. Overall detection levels are obviously lower than those of the total intensity, but detection levels of the kinematic component are still quite significant. Since this component is more spatially extended, it contributes to the overall S/N levels at lower multipoles, which explains the relatively constant detection level for the various redshifts when the detector noise is dominant, unlike the case of thermal effect detection. This also explains the sharp decrease in S/N levels at low redshifts, where contribution from lower multipoles is more dominant. As the redshift of the cluster increases, the signal becomes prominent on the noisy range of the multipoles until the redshift increases above $\sim 0.04$ and S/N levels rise, leveling off at $z \gtrsim 0.2$.

\begin{figure*}
\centering\includegraphics[width=\textwidth]{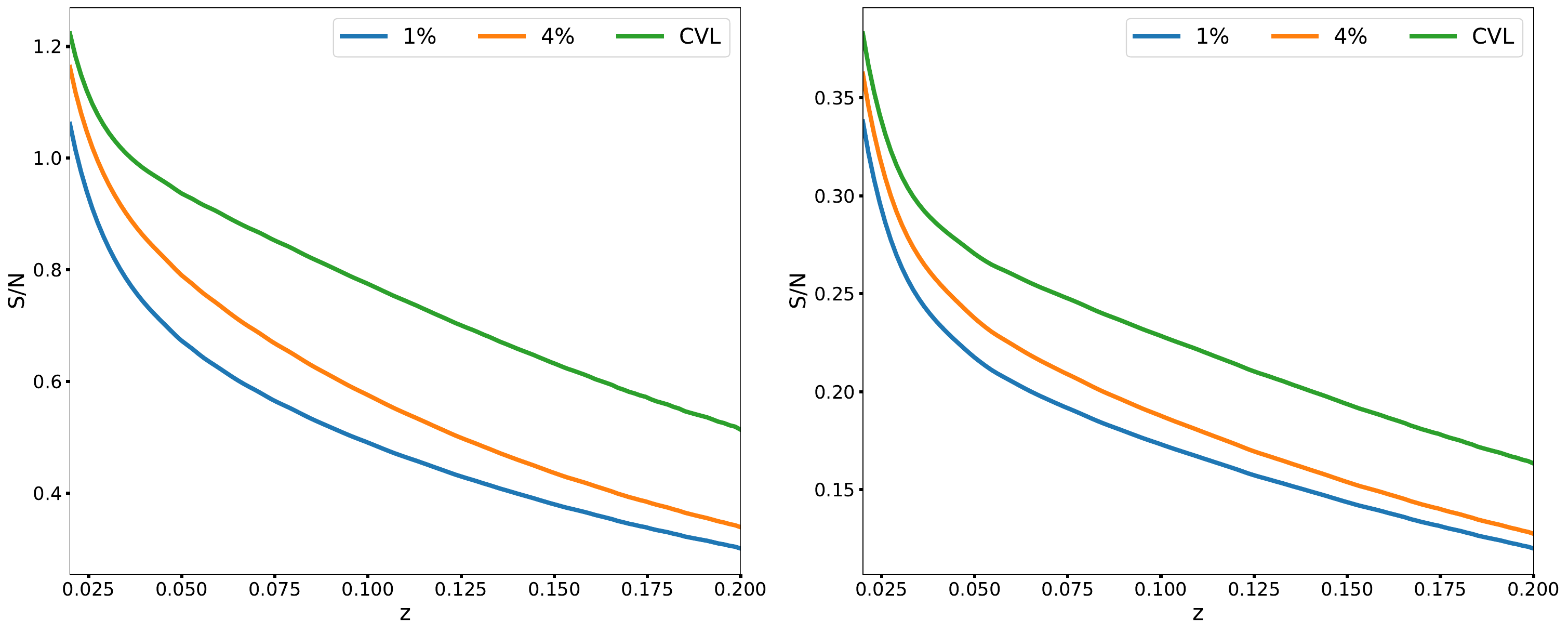}
\caption{Detection likelihood of the $\propto\tau^{2}\Theta$ (left) and $\propto\tau^{2}\beta_{t}$ (right) polarized signals from the model cluster for an SO-like telescope. Observation times and corresponding line styles are the same as in Fig.~\ref{fig:S2N_Temperature}. Point source noise contamination is assumed at the deduced upper limit (specified in the text).}
\label{fig:S2N_Polarization}
\end{figure*}

Estimates of S/N levels of the thermal ($\propto\tau^{2}\Theta$) and kinematic ($\propto\tau^{2}\beta_{t}$) SZ polarization components that are induced by double-scattering toward the model cluster, when observed for the specified integration times, are plotted as a function of redshift in the left- and right-hand panels, respectively, of Fig.~\ref{fig:S2N_Polarization}. In these estimates, the level of polarized point source emission is taken to be the upper limit mentioned earlier in this section. All other polarization components from the same cluster have different spatial and frequency dependencies and are therefore included in the noise budget for the detection of the $\propto\tau^{2}\Theta$ component. As mentioned in Sect. 3, their contributions to the polarization noise budget are negligibly small compared to the primordial contributions and are therefore neglected in our analysis. We have explicitly verified that the maps of the four polarization components explored in the present work only weakly correlate with one another. Since $\beta_{t}$ is assumed to be constant, the $\propto\tau Q_{{\rm prim}}$ and $\tau\beta_{t}^{2}$ maps are fully spatially correlated. Although they have different frequency dependencies, as specified in Sect.~\ref{sec:polarization_induced_by_scattering in_clusters}, they are nevertheless strongly correlated over the frequency range covered by the SO. These polarization components are correlated with the $\propto\tau^{2}\beta_{t}$ and $\propto\tau^{2}\Theta$ components at the 5\% and 0.3\% levels, respectively. The $\propto\tau^{2}\beta_{t}$ and $\propto\tau^{2}\Theta$ are correlated at the 6\% level.

\begin{figure*}
\centering\includegraphics[width=\textwidth]{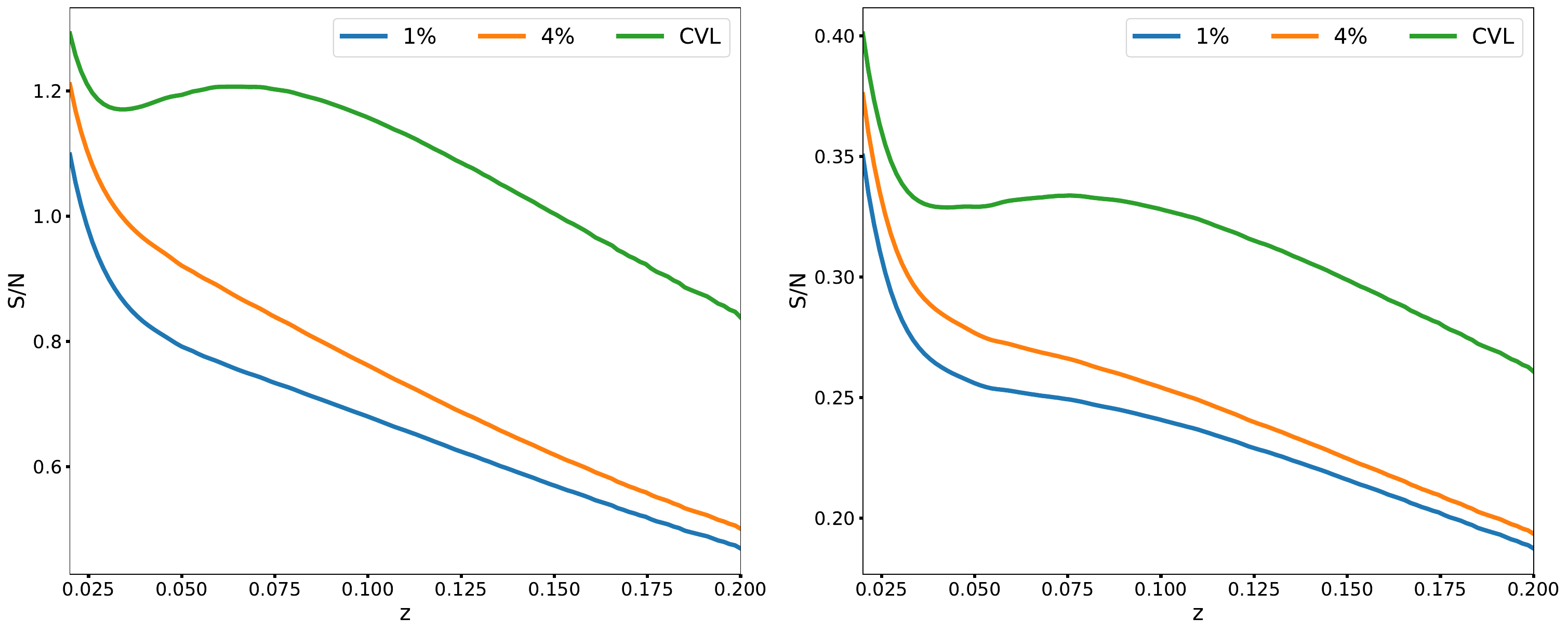}
\caption{Same as Figure~\ref{fig:S2N_Polarization} but with one-third of the currently deduced upper limit on point source contamination.}
\label{fig:S2N_Polarization_0p3}
\end{figure*}

As expected, the measurement of cluster polarized signals is challenging, with only marginal detection significance of clusters at $z\lesssim0.04$ at the specified observation times. Consequently, detection of polarized cluster-induced signals from individual rich clusters is quite unlikely. However, the effect could be statistically detected in observations of several carefully selected clusters. Point source emission and detector noise contribute most to the detection uncertainty. For longer observation times, point source contamination becomes more prominent at high multipoles (mostly affecting clusters at high redshift), which results in a relatively slow S/N falloff with redshift due to the constant (in $\ell$-space) point source contamination. The overall shape of the polarization S/N curves is different than those of the respective total intensity curves due to the relative contributions of primary CMB, point source, and instrument noise compared to the integrated thermal SZ contribution to the noise budget of total intensity and their different dependences on the angular scale. The impact of these sources of noise is much stronger on the polarization S/N curves, especially for more distant clusters, so much so that S/N values monotonically decrease (rather than increase) with redshift. In other words, the respective S/N curves would have been similar in shape had the polarized point source contribution been appreciably lower than the value assumed in our calculations.

To determine the dependence of the S/N estimates on a realistic level of uncertainty in key parameters, we repeated the calculations assuming an uncertainty of $\pm10\%$ in the values of the central temperature and density, and $\pm20\%$ in the value of $\beta$. The assumed uncertainty in the first two quantities results in up to $\mp17\%$ in the predicted values of the polarization S/N. For a value of $\beta$ lower by $20\%$ than $2/3$, the polarization S/N are too low for detection, whereas for (the $20\%$ higher value) $\beta \simeq 0.8$, the detection statistical significance reaches just over $3\sigma$ for $z\lesssim 0.02$. The assumed parameter uncertainties do not appreciably affect the likelihood of detection of the kinematic SZ effect.

It is also of interest to assess the likelihood of detecting SZ polarization via cross-correlation of the polarization and total intensity signals. Our numerical calculations show that the cross-correlation term, $-2{\rm Re}(T_{{\ell},(\nu_1}^{*}E_{{\ell},\nu_2)})C_{\ell}^{TE}$, in Eq.~\ref{eq:s2n_T_P} has only a very small impact on the S/N for polarization detection, and therefore no extra information can be deduced from it on the cluster-induced polarization of the CMB.

The incomplete sky coverage considered in this work is determined by the FOV of the telescope. As is well known, the non-local E and B polarization modes are sensitive to the nontrivial boundary conditions of the observed sky; mixing between $E_{\ell}$ and $B_{\ell'}$, where $\ell\neq\ell'$, takes place over a wide range of multipoles roughly determined by $\Delta\ell\sim\pi/\Delta\theta$, where $\Delta\theta$ is the angular size of the observed sky. In fact, the degree of mixing depends not only on the angular area of the observed sky patch, but also on its geometrical shape. To test for the possible impact of an alias of power between the E- and B-modes due to the incomplete sky coverage on our analysis, we considered synthetic maps that would ideally be pure E-mode on the full sky. Observing these maps over our FOV generates some E-to-B leakage due to the nontrivial boundaries of the cut sky. It is clear from Eq.~\ref{eq:s2n_T_P} that in order to estimate the degree of E-B mixing, the relevant quantities to be compared are $\mathcal{P}_{E}=\int \frac{d^{2}{\ell}}{(2\pi)^{2}}|E_{\ell}|^{2}$ and $\mathcal{P}_{B}=\int \frac{d^{2}{\ell}}{(2\pi)^{2}}|B_{\ell}|^{2}$, which represent the total power of the E and B maps, respectively, where the integration extends over all wave-vectors $\bf{\ell}$. The ratio $\mathcal{P}_{B}/\mathcal{P}_{E}$ computed for our (ideally pure E-mode) maps is a reasonably good measure of the E-B mixing due to the cut sky. Our numerical estimates indicate that $\mathcal{P}_{B}/\mathcal{P}_{E}$ never exceeds 0.1, implying that the E-B mixing level is at the $\lesssim 10\%$ level, which is, at worst, comparable in significance to the other inherent uncertainties in our cluster modeling, as described above (which ultimately propagate into our estimation of signal detectability).

Since the point source contamination affects the feasibility of detecting SZ polarization toward clusters, and given the fact that the current upper limit may well be too conservative, it is only reasonable to gauge the improved detection likelihood when its level is assumed to be appreciably lower. To do so, we recalculated S/N values with point source polarized intensity contamination taken to be, for example, one-third of the deduced upper limit specified earlier in this section; results of these calculations for the thermal SZ polarization are shown in Fig.~\ref{fig:S2N_Polarization_0p3}. As is evident from the figure, the impact of this lower level of polarized point source signal is a somewhat higher level of S/N and a less steep decrease with redshift, but the likelihood of detecting cluster-induced polarization is still quite low. As mentioned previously, with lower point source power and no detector noise (the CVL case in Fig.~\ref{fig:S2N_Polarization_0p3}), a slight increase in S/N around $z\sim0.05$ is apparent; this resembles the behavior of the S/N curves for intensity detection.

Although the detection of cluster-induced polarization toward individual clusters will be quite challenging, it could nevertheless be detected statistically: As is clear from Figs.~\ref{fig:S2N_Polarization}-\ref{fig:S2N_Polarization_0p3}, increasing the observation time from 1\% to 4\% of the nominal observation time results only in a small boost of the S/N for its detection toward individual clusters. A more feasible strategy would be the observation of the effect toward a few clusters, while spending only, for example, 1\% on each cluster. From the S/N levels shown, it is evident that a few months of "shallow" observations toward $\sim 20$ individual clusters could result in a $3\sigma$ statistical detection of the effect. In addition, observation time can be optimized by specifically selecting sky patches with an especially rich cluster population.

\section{Discussion}

There is considerable interest in measuring the polarization of individual super-galactic systems in order to determine the spectral and spatial properties of the dominant sources that constitute the inherent confusing foreground in CMB measurements. Reasonably detailed knowledge of microwave foregrounds is needed for a correct identification of any residual CMB signals. As the largest bound systems, galaxy clusters are the most important contributors to the foreground on scales larger than a few arcminutes. This further enhances the intrinsic interest in spectral and spatial SZ mapping of individual clusters, which is diagnostically important for determining basic cluster properties: the IC gas pressure profile, the gas and total mass of the cluster, and the cluster velocity. The transverse velocity component of the cluster can only be determined from the measurement of the kinematic polarization signals described in Sect.~\ref{sec:polarization_induced_by_scattering in_clusters}.

We calculated the total and polarized intensities resulting from scattering in a fiducial rich cluster whose total mass is $2\times 10^{15} \,M_{\odot}$ with polytropic IC gas whose mass fraction is $f_{g}=0.12$; the cluster is assumed to have radial and tangential velocity components of $1000$ km s$^{-1}$. For a fiducial SO-like CMB polarization project, we calculated the predicted S/N of the thermal and kinematic total and polarized SZ intensities as function of the cluster redshift. Given the high measurement sensitivity of the experiment, it is only to be expected that the mapping of the thermal SZ component across nearby clusters would be readily accomplished. Dedicated long measurements with ground-based telescope arrays will make it possible to determine the gas pressure across most of the cluster extent, not just the central region. This will yield the gas density profile and, consequently, the total mass profile, thereby resulting in a more precise determination of the cluster mass within the virial radius.

More demanding but still quite likely is the projected capability to measure the kinematic SZ component toward fast moving clusters. As is apparent from the left-hand panel of Fig.~\ref{fig:S2N_Temperature}, observing a cluster moving radially with a velocity of $\sim 1000$ km s$^{-1}$ for more than $\sim 1\%$ of their respective observing time will result in a very significant detection of the kinematic SZ component. Significant detection of this component is also likely for slower moving clusters (when observed for the same times). For example, at $\sim 500$ km s$^{-1}$, detection at $S/N > 3$ is likely for clusters at redshifts in the full range considered here ($z \leq 0.2$). This indicates that the radial velocity component of nearby clusters could be measured at a high level of precision by future SO-like ground-based CMB experiments.

In this study, the fiducial rich cluster was taken to be spherically symmetric, relaxed, and in hydrostatic equilibrium. Clearly, the predicted level of induced polarization depends very much on the actual gas morphology and its velocity field, in addition to the temperature and density profiles. The former two characteristics can be very different in a merging cluster from those in a relaxed cluster, so much so that the feasibility of detection of the kinematic and total SZ polarization signals could be significantly higher than estimated here. Whereas a simple scaling of the kinematic polarization signals toward a merging cluster based on the possibly higher value of the transverse velocity component, as compared to the value assumed here ($\sim 1000$ km s$^{-1}$), would seem to be adequate for estimating the enhanced kinematic polarization signals, a more realistic analysis would be needed in order to account for the non-spherical distribution of the gas and the spatial profile of the velocity field. This can be reliably accomplished based on the detailed mapping of these quantities by a high resolution hydrodynamical simulation of a merging cluster.

The statistical cluster polarization signal is very small, as is apparent from the predicted polarization power spectra calculated by, for example, \citet{2006MNRAS.368..511S}. Ground-based observations aimed at detecting the predicted inflationary B-mode polarization typically target only few percent of the sky. Even though these (radio-quiet) patches are optimally chosen, polarization signals induced by individual clusters can still have an overall effect of more than the conservative level of a few percent residual contamination after subtraction of the statistical cluster signal estimated in \citet{2006MNRAS.368..511S}. In principle, the cluster polarization power spectra could also affect the precision of global parameter estimation, but these are too weak to impact the overall error in deduced parameter values (\citealt{Cooray:2004cd}, \citealt{2009MNRAS.399.2088S}). A possible exception could be their effect on the residual lensing-induced B-mode signal, which can amount to a few percent, depending on the accuracy with which the lensing signal can be removed.

\begin{acknowledgements}
We thank the referee for useful comments and suggestions made on an earlier version of the manuscript. This work has been supported by grants from the Israel Science Foundation, and the Joan and Irwin Jacobs donor-advised fund at the Jewish Community Foundation (San Diego, CA). MM acknowledges support from the European Research Council under the European Union's Seventh Framework Programme (FP/2007-2013) / ERC Grant Agreement No. [616170]. This research used resources of the National Energy Research Scientific Computing Center (NERSC), a U.S. Department of Energy Office of Science User Facility operated under Contract No. DE-AC02-05CH11231.
\end{acknowledgements}

\bibliographystyle{aa}
\bibliography{pol_clus_bib}

\appendix
\section{Polarization maps}
\label{appendix_A}

\begin{figure}[!h]
\begin{center}
\subfigure[]{\includegraphics[width=4.3cm,trim={0  2.4cm 0 0}]{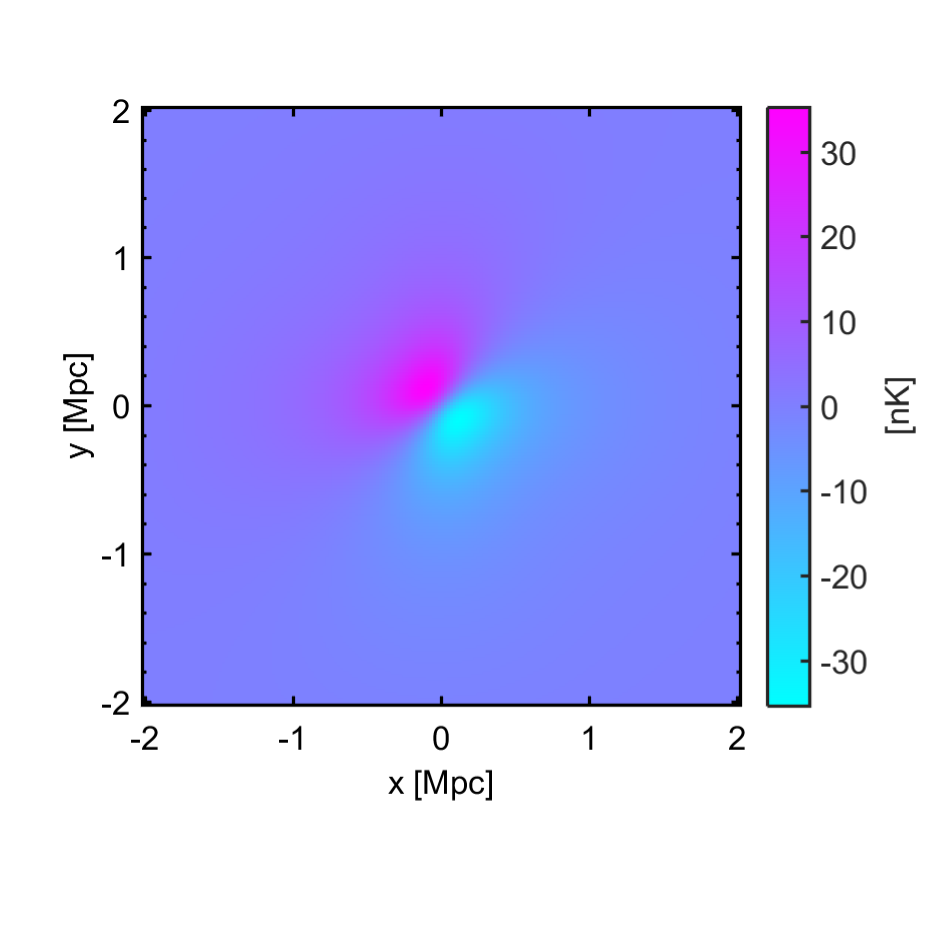} \label{fig:Q2}}
\subfigure[]{\includegraphics[width=4.3cm,trim={0  2.4cm 0 0}]{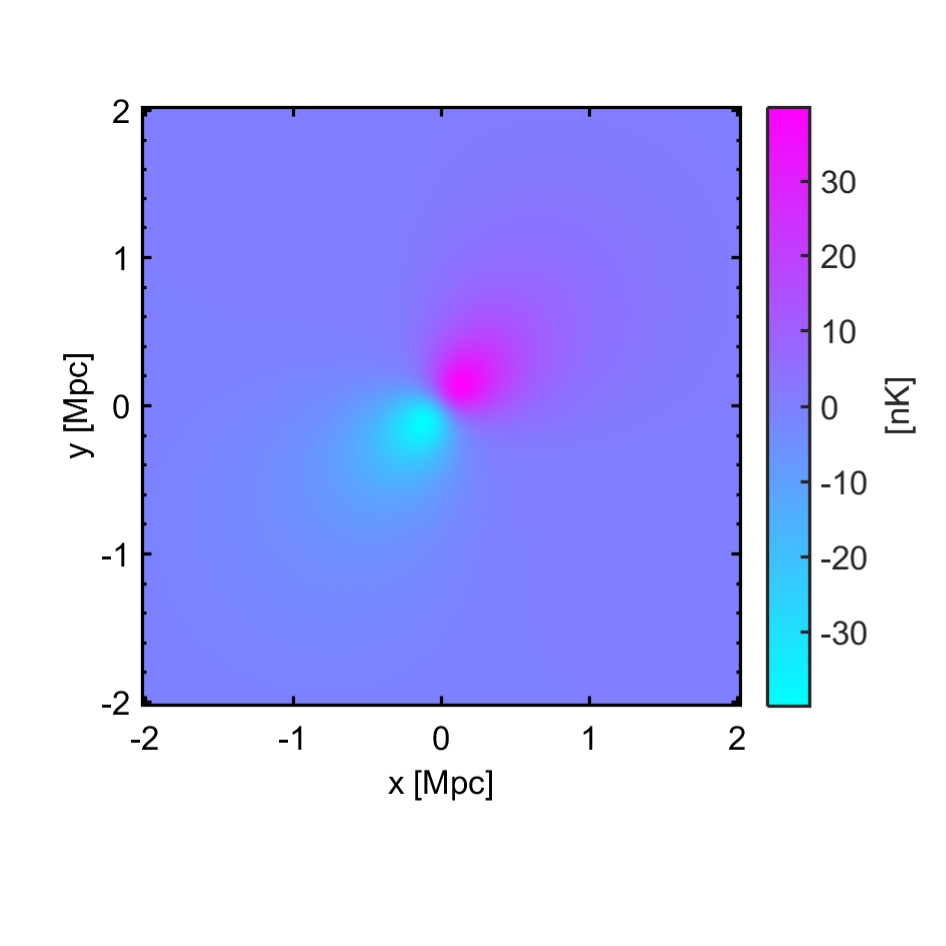} \label{fig:U2}}  \\ \vspace*{-4mm}
\subfigure[]{\includegraphics[width=4.3cm,trim={0  2.4cm 0 0}]{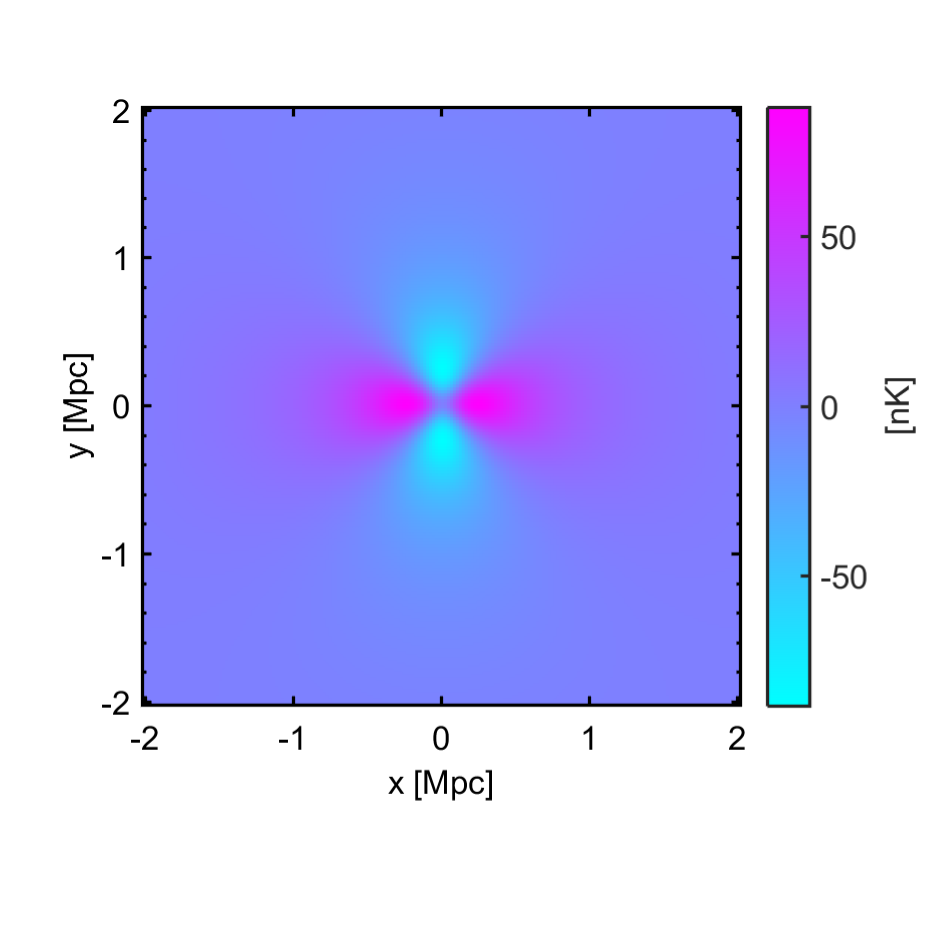} \label{fig:Q3}}
\subfigure[]{\includegraphics[width=4.3cm,trim={0  2.4cm 0 0}]{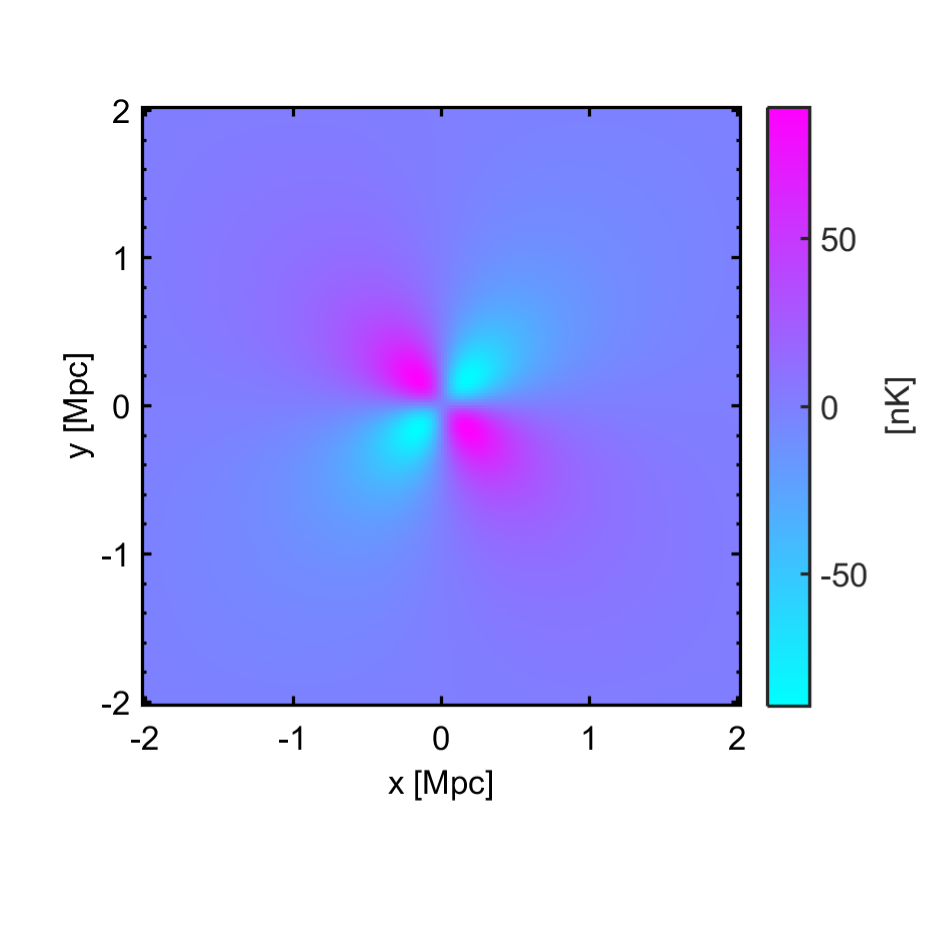} \label{fig:U3}}
\end{center}
\caption{Real-space Q (left) and U (right) maps of the two double-scattering polarization effects ($\tau^2\beta_{t}$ in \subref{fig:Q2}-\subref{fig:U2} and $\tau^2\Theta$ in \subref{fig:Q3}-\subref{fig:U3}). The latter effect is shown here for $\nu=150$ GHz and scales to other frequencies according to the first line of Eq.~\ref{eq:f_x_def}. The two other polarization terms (linear in $\tau$) have a trivial pattern and were therefore not presented here.}
\label{fig:pol_maps}
\end{figure}

\end{document}